\documentclass[aps,prl,twocolumn,groupedaddress,amsmath,floatfix,showpacs]{revtex4}
\usepackage{graphics,graphicx,epsfig}
\begin{document}

\title{Dielectric Response of Periodic Systems from Quantum Monte Carlo Calculations}
\author{P. Umari,$^{1}$ A. J. Willamson,$^2$ 
Giulia Galli,$^2$ and Nicola Marzari$^1$}
\affiliation{$^1$Department of Materials Science and Engineering, MIT, 02139 Cambridge MA}
\affiliation{$^2$Lawrence Livermore National Laboratory, 94550 Livermore CA}

\date{\today}

\begin{abstract}
We present a novel approach that allows to calculate the dielectric response
of periodic systems in the quantum Monte Carlo formalism.
We employ a many-body generalization for the electric enthalpy functional,
where the coupling with the field is expressed via the
Berry-phase formulation for the macroscopic polarization.
A self-consistent local Hamiltonian then determines the ground-state 
wavefunction, allowing for accurate diffusion quantum Monte Carlo 
calculations where the polarization's fixed point is estimated from the average 
on an iterative sequence, sampled via forward-walking.
This approach has been validated for the
case of an isolated hydrogen atom, and then applied to a periodic system, to
calculate the dielectric susceptibility of molecular-hydrogen chains.
The results found are in excellent agreement with the best estimates
obtained from the extrapolation of quantum-chemistry
calculations. 
\end{abstract}


\maketitle

The response of an extended system to an electric field is
an intrinsic property of central importance to the understanding
and characterization of bulk materials.
First, it depends very sensitively on an accurate description of
the correlations between interacting electrons.
Second, it allows one to predict observable quantities such as the infrared 
or (non-resonant) Raman and hyper-Raman spectra, establishing 
a direct link between macroscopic experimental observations and
microscopic calculations.
Density-functional approaches to calculate
dielectric susceptibilities in periodic systems were introduced more than two
decades ago \cite{km82,br86,hl87,bgt87}.
However, the resulting agreement with experiments is less striking
than for the case of structural or vibrational properties, 
even though the response to a static electric field is strictly
a ground-state property \cite{br86}. 
These discrepancies have been related to the intrinsic dependence of the exchange and
correlation functional on the polarization \cite{ggg95};
non-local exchange-correlation functionals
such as the weighted-density approximation (WDA)
\cite{ms00} also show some improvements upon the local-density (LDA) \cite{br86} or generalized-gradient
approximations (GGAs) \cite{dbr94}.
While in the case of crystalline semiconductors dielectric susceptibilities are usually 
overestimated by $10-30$\% in LDA or GGAs, 
linear dielectric susceptibilities of conjugated polymers
are usually overestimated by a factor $>2-3$, and 
non-linear susceptibilities by orders of magnitude \cite{cpg98,gsg99,fbl02}.
Linear chains of hydrogen dimers are another clear example of LDA or GGAs failures, 
and have thus become a stringent test case to assess 
new developments in density-functional theory
\cite{fbl02,mwy03,kkp04,bn05}.
Correlated quantum-chemistry methods are able to
predict accurate linear and non-linear polarizabilities for  
polymeric chains \cite{ttm95}, but their
less favourable scaling and 
the need, in the absence of periodic boundary-conditions (PBCs), to extrapolate the results to
the infinite system limit their range of applicability.

Highly-accurate ground-state electronic-structure calculations can be performed using
continuous quantum Monte Carlo (QMC) methods \cite{fmn01}, which benefit from scaling costs
as low as $N-N^3$ \cite{whg01}, where $N$ is the number of electrons.
Although diffusion QMC has been successfully applied to address the polarizability of
small molecules \cite{sf99}, the extension to bulk materials
is hindered by the difficulty
in treating the response to a linear electric field,
since the latter is non-periodic and incompatible with PBCs.
Only the response of the electron gas to a finite wavelength perturbation
has been investigated \cite{mcs92}.
In density-functional calculations, the response to an electric field has been calculated
via its long-wavelength limit using supercells of increasing size \cite{m83,km82,ms00},
or, more elegantly, using linear-response theory \cite{bgt87,gv89}. 
A general approach able to 
deal with finite fields, and thus both linear and non-linear effects, has also
been recently introduced \cite{up02,siv02}.
This method is based on the fact that the macroscopic polarization can be properly defined 
even in PBCs \cite{kv93,r92}; then, instead of applying a field in the form a linear potential,
a Legendre transform allows one to introduce an electric enthalpy functional whose minimization
leads to the correct wavefunctions.

In this work we apply these ideas to the case of correlated electrons,
using the many-body formulation of the Berry-phase polarization \cite{r98,wm01}
to deal with the generalization of the electric-enthalpy functional 
\cite{nv94,ng01,up02,siv02} to the interacting case \cite{sbp04}.
We show how such an approach can be applied to the case of diffusion quantum Monte Carlo (DMC) 
calculations, where a local Hermitian operator is needed to evolve a population 
of walkers representing the ground state.

Let's consider a system of $N$ interacting electrons in a periodic
cell of size $L$ (to simplify the notation we describe the
one-dimensional case, but the extension to higher dimensions is
straightforward). 
The normalized many-body wavefunction $\Psi$ obeys  PBCs
\begin{equation}
\Psi(x_1,..,x_i+L,..,x_N)=\Psi(x_1,..,x_i,..,x_N).
\end{equation}
The polarization of the system can be obtained from 
the single-point Berry phase \cite{r98}
\begin{eqnarray}
\label{pola}
P[\Psi]&=&-\frac{e}{\Omega}\frac{L}{2\pi} {\mathrm{ Im}} \ln z\\
\label{zeta}
z[\Psi]&=&\langle\Psi|e^{iG\hat{X}}|\Psi\rangle,
\end{eqnarray}
where $\Omega$ is the size of the cell \cite{notaomega}, $e$ is the electron charge, $G=2\pi/L$, and 
$\hat{X}$ the $N$-body operator
\begin{equation}
\hat{X}=x_1+x_2+...+x_N.
\end{equation}
The definition of Eq.\ (\ref{pola}) coincides, in the thermodynamic limit,
with the exact many-body observable \cite{r98}, but it also remains well-defined for
any finite $L$. 
With the  many-body polarization well defined,
the ground-state wavefunction of a periodic, extended system
in the presence of an electric field $\mathcal{E}$ 
can be obtained from the minimum of the generalized electric enthalpy 
\cite{nv94,ng01,up02,siv02,sbp04} 
\begin{equation}
\label{entalpia}
F[\Psi]=E^0[\Psi]-{\mathcal E}\Omega P[\Psi],
\end{equation} 
where $E^0[\Psi]$ is the energy functional for the unperturbed Hamiltonian $H^{0}$.

The functional in Eq.\ (\ref{entalpia}) could be directly minimized using a variational
Monte Carlo approach \cite{footnote}, but it can't be applied directly to the 
more accurate case of DMC calculations, since these can only deal with Hamiltonians
that are both local and Hermitian.  A local Hermitian operator can nevertheless be derived
from the minimum condition for the electric enthalpy
\begin{equation}
\frac{\delta F}{\delta\langle\Psi|}=\lambda|\Psi\rangle ,
\end{equation}
with $\lambda$  an appropriate Lagrange multiplier \cite{notabene}.
The $\Psi$ that minimizes Eq.\ (\ref{entalpia})
is also the ground-state for the {\it many-body self-consistent Hamiltonian}
\begin{equation}
\label{hamilt}
H(z)=H^{0}+{\mathcal E}\frac{eL}{2\pi}
{\mathrm{Im}}\frac{e^{iG\hat{X}}}{z}
\end{equation}
where $z$ depends on $\Psi$ through Eq.\ (\ref{zeta}).
For any given $L$, there exists a well-defined interval for ${\mathcal E}$
centered around $0$
for which the electric enthalpy can be minimized without encountering
runaway solutions \cite{up02,siv02}.
In addition, all physical observables of interest here
(e.g. the linear and nonlinear dielectric susceptibilities) are
derivatives of the electric enthalpy with respect to the 
field and thus remain well-defined for every $L$.

Due to the self-consistent nature of the operator $H(z)$ defined in Eq.\ (\ref{hamilt}),
the ground state in the presence of an electric field must be found through  
an iterative procedure. We start from a first value $z_1$ for $z$, e.g. as found in
the single-particle calculations or in the many-body trial wave function $\Phi_{\mathrm T}$; 
the local Hamiltonian $H(z_1)$ is then constructed. DMC evolution using this operator leads
to a new expectation value for $z$, called
$z_2$, which in turn determines a second Hamiltonian $H(z_2)$.
In the absence of stochastic noise, this process could be iterated to convergence
$z_1\rightarrow z_2\rightarrow z_3\rightarrow...\rightarrow z_n$,
to find the fixed point of the complex-plane map
\begin{equation}
f(z_i)=z_{i+1}.
\end{equation}
Since the Monte Carlo procedure introduces a statistical error 
in every estimate of $z_i$, 
the map $f$ becomes a stochastic function in the complex plane.
Now, the linear approximation $f(z)=a z + b$ 
is made close to the fixed point of $f(z)$
(linearity will be shown numerically later). Then,
the {\it average over a sequence of $\{z_i\}$} provides the estimate for the 
fixed point $b/(1-a)$ of $f(z)$,
from which the polarization is then obtained via Eq.\ (\ref{pola}). 
This descends straightforwardly from the consideration that
$\lim_{N\rightarrow\infty} \frac{1}{N}\sum_{i=1,N} z_i$ 
is the fixed point of the map $f(z)$, 
since $\frac{1}{N}\sum_i z_i=\frac{1}{N}\sum_i f(z_{i-1})=\frac{1}{N}\sum_i f(z_{i})=\frac{1}{N}\sum_i 
(a z_i+b)$ (the second equality is valid in the limit ${N\rightarrow\infty}$). 
Finally, supposing that the error in the complex plane is 
isotropic with a quadratic spread $\sigma^2$, the \{${z_i}$\} will be distributed
around the fixed point with a quadratic spread $\sigma^2/(1-|a|^2)$.

Most implementations of DMC use
importance sampling: during diffusion in imaginary time, 
the walkers at a time $t$ 
are distributed as $\Phi_{\rm T}({\bf R})\Psi({\bf R})$,
where $\Psi$ is the correct ground-state wavefunction and $\Phi_{\rm T}$ is the
trial wavefunction ($\Phi_{\rm T}$
is chosen here to be the product of a Jastrow factor and a Slater determinant).
Since the operator $e^{iG{\hat X}}$ does not
commute with the Hamiltonian, we use a forward-walking strategy \cite{lkc74,hlr94} 
to sample the expectation value $z$.
In a branching DMC approach
the estimate of  $\Psi({\bf R}_{j,t})/\Phi_{\rm T}({\bf R}_{j,t})$
for any walker $j$ at a time $t$ and position ${\bf R}_{j,t}$ is
given by the number of its descendants after a
projection time $\Delta t$.
The expectation value $z$ of $e^{iG{\hat X}}$ then becomes
\begin{equation}
\label{forward}
{\overline{ e^{iG{\hat X}}}}=\frac{\sum_\tau\sum_{j=1,N_\tau} e^{iGX'_{j,\tau-\Delta t}}}
{\sum_\tau N_\tau},
\end{equation}
where $N_\tau$ is the number of walkers at timestep $\tau$,
and $X'_{j,\tau-\Delta t}$ corresponds
to the ascendant configuration of the walker 
 $j$ at an interval $\Delta t$ back in imaginary time.

\begin{table}
\begin{tabular}{lccc}
Method&10-H$_2$&16-H$_2$&22-H$_2$\\ \hline
DFT-GGA&102.0&123.4&133.5\\
DMC & 52.2 $\pm$ 1.3  & 55.4 $\pm$ 1.2 & 53.4 $\pm$ 1.1
\end{tabular}
\caption{\label{tab1}
Linear polarizability per H$_2$ unit for a periodic linear chain
of  H$_2$ dimers, calculated in PBCs for supercells containing 10, 16, and 22
H$_2$ units.}
\end{table}

\begin{table}
\begin{tabular}{lcc}
Method    &  $\alpha$ (a.u.)   & Scaling cost \\ \hline
PBE-GGA   &   144.5              & N-N$^3$\\ \hline
CCD       & 47.6  & N$^5$-N$^6$\\
CCSD    & 48.0  & N$^6$\\
CCSD(T)   & 50.6  & N$^7$\\\hline
MP2       & 58.0  & N$^5$\\ 
MP3       & 54.3  & N$^6$\\
MP4       & 53.6  & N$^7$\\\hline
DMC       & 53.4$\pm$ 1.1  &   N-N$^3$
\end{tabular}
\caption{\label{tab2}
Linear polarizability per H$_2$ unit for a periodic linear chain
of  H$_2$ dimers calculated using several methods, together
with their associated scaling costs.
The quantum chemistry results are from Ref.\ \protect\cite{cmv95}.}
\end{table}

We implemented this approach in the QMC {\sc Casino} package \cite{casino};
the single-particle orbitals for the Slater determinant of the trial wavefunction
$\Phi_{\mathrm T}$ were obtained from {\sc Quantum ESPRESSO} \cite{pwscf},
using consistently the same finite electric field ${\mathcal E}$ of the 
QMC calculation \cite{espresso}.

\begin{figure}
\begin{center}
\includegraphics[width=7.0cm]{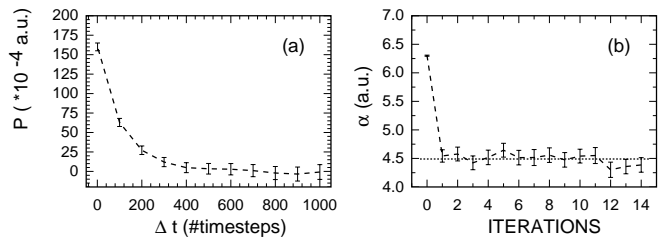}
\end{center}
\caption{\label{fig1}
(a) Polarization of an isolated H atom 
{\it in the absence} of an electric field as a function of the
forward walking parameter $\Delta t$,
using a trial wavefunction  expressed by a PBE-GGA orbital 
calculated {\it in the presence} of  an electric field of 0.005 a.u.\ .
The dashed line is a guide to the eye.
(b) Polarizability of an isolated H atom, 
for each of the
self-consistent DMC runs in the iterative sequence (see text).
The dashed line is a guide to the eye.
}
\end{figure}

To validate our approach, we first calculate the polarizability of an
isolated hydrogen atom, using a large supercell and periodic-boundary conditions.
For this case, a homogeneous finite field can be explicitly applied using a
saw-tooth potential (i.e. a potential that is both periodic and piecewise linear), and
the results compared with those obtained with our method.
We use the same GGA trial wavefunctions for both tests, as calculated
under an applied  electric field of $0.005$ a.u. using a
local H pseudopotential, $50$ Ry cutoff in the plane-wave basis set, and
a cell of $20 \times 20 \times 200$ a. u. (the longest direction coincides with the
applied electric field, and a large supercell has been used to eliminate finite-size
errors \cite{up03}).
In all DMC calculations we
choose a timestep of $0.02$ a.u., 
assuring an acceptance ratio greater than 99.7\%.
Then, the 
projection time $\Delta t$ for the forward-walking 
is determined.  We show in Fig.\ \ref{fig1}(a)
the dependence of the expectation value for the polarization 
of the atom in the absence of an applied field, as a function of 
the projection time $\Delta t$ in Eq. (\ref{forward}), showing convergence
to  $0$ a.u.  for
$\Delta t \geq 600$ timesteps.
With these parameters, we start our iterative sequence of calculations.
The first value for $z$ is taken from the GGA results, and then after each 
DMC run the local operator $H(z)$ is updated using the current expectation
value for $z$.
We report in Fig.\ \ref{fig1}(b) the 
atomic polarizability calculated from each one of the $z_i$  
along the sequence of DMC runs.
The average of the $z_i$ over the last $13$ runs
gives us a  fixed point estimate $(0.9995072\pm 2.*10^{-7} +7.05*10^{-4}i\pm4.*10^{-6}i)$ for $f(z)$, and a corresponding 
polarizability for the H atom of $4.49 \pm 0.03$ a.u., in
excellent agreement with the result obtained for the saw-tooth potential
($4.52 \pm 0.05$ a.u.) and the exact value of $4.5$ a.u. \cite{w26}.

\begin{figure}
\begin{center}
\includegraphics[width=7.0cm]{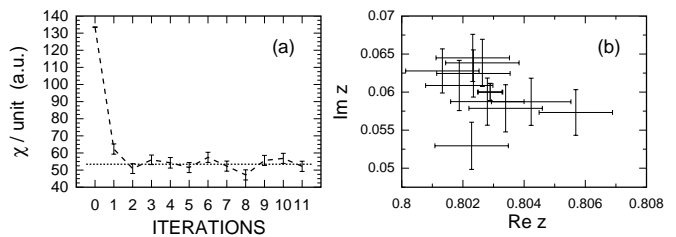}
\end{center}
\caption{\label{fig2}
(a) Linear polarizability per H$_2$ unit for a periodic linear chain
of  22 H$_2$ dimers in PBCs,
for each of the self-consistent DMC runs in the iterative sequence (see text).
The dashed line is a guide to the eye.
(b) Last ten $z_i$ values (thin), and estimate of the fixed point
   of $f(z)$ (bold).
}
\end{figure}

Having validated our scheme in an isolated case, we proceed to 
apply it to a genuine periodic system, namely  a linear chain of hydrogen dimers, where
standard density-functional theory performs very poorly \cite{fbl02}.
We adopt one of the standard geometries
reported in the literature \cite{cmv95,gsg99},
where the hydrogen dimers
have a H-H bond length of 2.0 a.u. and are separated by 2.5 a.u. .
For this case, reliable reference results for
the polarizability per dimer unit $\alpha$ are available from
correlated quantum-chemistry approaches  ($\alpha=53.6$ a.u. in MP4 and
$\alpha=50.6 $ a.u. in CCSD(T)) \cite{cmv95};
these numbers should be contrasted with the PBE-GGA result $\alpha=144.5$ a.u. \cite{cutoff}.
To calculate the DMC susceptibility, we use a trial wavefunction composed of a Slater determinant
of PBE-GGA single-particle orbitals (consistently calculated with an electric field of $0.001$ 
a.u.) times a Jastrow term containing the
electron-electron and electron-nucleus terms \cite{simmetria}.
We consider three supercells containing $10$, $16$ and $22$ H$_2$ units,
and the same electric field of $0.001$ a.u. .
The polarizabilities per unit dimer are  found by finite differences 
(note that the  polarization at zero field is zero by symmetry).
In Fig.\ \ref{fig2}(a), we show the polarizability calculated at each
DMC run in the 22-unit supercell. 
After the second iteration, the instantaneous polarizability oscillates roughly around the final value;  again, we stress 
that it is the average of \{$z_i$\} that delivers the estimate of the fixed point and thus
the final polarization.
In Fig.\ \ref{fig2}(b), we show the distribution of the last ten $z_i$,
and  the fixed point  $( 0.8029\pm 4.*10^{-4} + 6.0*10^{-2}i \pm  1.*10^{-3}i)$,
obtained from their average.
In Table \ref{tab1}, we illustrate the convergence of the calculated polarizability
with respect to the supercell size \cite{notabene2}:
The faster convergence with respect to the PBE-GGA case originates in the
the stronger localization  of $\Psi$ in DMC.
Indeed, the localization spread $\lambda^2$ (defined as $-(L^2/N4\pi^2)\ln|z|^2$ \cite{rs99})
decreases from $4.32$ a.u. (i.e. Bohr$^2$) for PBE-GGA to $2.44 \pm 0.01$ a.u.\ in DMC
for the 22 units supercell.
Finally, we compare in Table \ref{tab2} our final result  $\alpha=53.4\pm1.1$ a.u.  with those
obtained using other approaches:  we find excellent agreement with the most accurate
results available using MP3 and MP4.
We note in passing that the correlated quantum chemistry polarizabilities have been extrapolated
from calculations performed for finite chains \cite{cmv95}, with the extrapolative correction amounting to
20\% of the final results.

In conclusion, we introduced and validated a novel approach to 
study periodic systems in the presence of finite electric-fields with
highly-accurate, and in principle linear-scaling, diffusion QMC
calculations.  In particular, we presented the first QMC evaluation of the
dielectric susceptibility of an extended system.
For the paradigmatic test case of hydrogen chains, we obtained
excellent agreement with the best quantum-chemistry
results. Our approach opens the possibility to evaluate
linear and non-linear dielectric properties, effective charges, and
Raman and infrared coupling tensors using QMC calculations.

This research has been supported by the DARPA 
grant B529527.
We are especially grateful to Richard Needs and Mike Towler 
of the University of Cambridge (UK)
for making their  QMC code {\sc Casino} \cite{casino} available to us.
Part of this work (AW and GG) was performed under the auspices of the
U.S. Department of Energy by the University of California, Lawrence
Livermore National Laboratory under contract No. W-7405-Eng-48.

\bibliographystyle{PhysRev}

\begin{thebibliography}{999}
\bibitem{footnote} Unless variance minimization \cite{uww88} is used; if the case, a local
Hermitian operator is required, and the approach presented here for DMC should be used.
\bibitem{notabene} This strategy has also been used for the case 
of single-particle Slater determinants 
(Eq. (64) of Ref.\ \cite{siv04}) and model Hubbard Hamiltonians
(Eq. (8) of Ref.\ \cite{sbp04}) .
\bibitem{espresso}
We implemented in the {\sc Quantum ESPRESSO} package an algorithm
analogous to the tight-binding one of Ref.\ \cite{siv04}.
\bibitem{simmetria} These terms in the Jastrow factor are spherically symmetric in the present implementation. 
Thus, the addition and optimization of the Jastrow factor to a Slater determinant of polarized PBE-GGA orbitals
shows only a marginal improvement (around 5\% for the hydrogen chain) for
the expectation value of the polarization.
\bibitem{notabene2} The DMC results were obtained from $z$-averages over
$10-11$ runs, each of them with $256 000$ time-steps and an average population of $1000$ walkers.
We verified linearity in the complex map, by repeating the calculation for $10-H_2$ using $29$ runs of
$64 000$ timesteps, obtaining $53.0 \pm 1.4$ a.u. .
\end{thebibliography}

\end{document}